\documentclass[reprint,aps,pra,twocolumn,superscriptaddress,floatfix,longbibliography]{revtex4-1}

\usepackage[dvips]{graphicx}
\usepackage{latexsym}
\usepackage{amsmath}
\usepackage{amsthm}
\usepackage{amsfonts}
\usepackage{amssymb}
\usepackage{mathptmx}
\usepackage{subfigure}
\usepackage{mathrsfs}
\newcommand{\parallelsum}{\mathbin{\!/\mkern-5mu/\!}}
\usepackage{color}
\definecolor{darkblue}{rgb}{0,0.02,0.45}
\usepackage[colorlinks=true,citecolor=darkblue]{hyperref}

\usepackage[all]{hypcap}
\usepackage{gensymb}

\definecolor{cream}{RGB}{222,217,201}     

\begin{document}

\clearpage

\title{Noncollinear magnetic structure and magnetoelectric coupling in buckled honeycomb Co$_4$Nb$_2$O$_9$: A single crystal neutron diffraction study}

\author{Lei Ding}
\affiliation{Neutron Scattering Division, Oak Ridge National Laboratory, Oak Ridge, Tennessee 37831, USA }

\author{Minseong Lee}
\affiliation{National High Magnetic Field Laboratory, Florida State University, Tallahassee, FL 32306-4005, USA}
\affiliation{Department of Physics, Florida State University, Tallahassee, Florida 32306-3016, USA}

\author{Tao Hong}
\affiliation{Neutron Scattering Division, Oak Ridge National Laboratory, Oak Ridge, Tennessee 37831, USA }

\author{Zhiling~Dun}
\affiliation{Department of Physics and Astronomy, University of Tennessee, Knoxville, Tennessee 37996-1200, USA}

\author{Ryan~Sinclair}
\affiliation{Department of Physics and Astronomy, University of Tennessee, Knoxville, Tennessee 37996-1200, USA}

\author{Songxue Chi}
\affiliation{Neutron Scattering Division, Oak Ridge National Laboratory, Oak Ridge, Tennessee 37831, USA }

\author{Harish~K.~Agrawal}
\affiliation{Neutron Sciences Directorate, Oak Ridge National Laboratory, Oak Ridge, Tennessee 37831, USA }

\author{Eun Sang Choi}
\affiliation{National High Magnetic Field Laboratory, Florida State University, Tallahassee, FL 32306-4005, USA}

\author{Bryan~C.~Chakoumakos}
\affiliation{Neutron Scattering Division, Oak Ridge National Laboratory, Oak Ridge, Tennessee 37831, USA }

\author{Haidong~Zhou}
\affiliation{Department of Physics and Astronomy, University of Tennessee, Knoxville, Tennessee 37996-1200, USA}
\affiliation{National High Magnetic Field Laboratory, Florida State University, Tallahassee, FL 32306-4005, USA}

\author{Huibo Cao}
\email{caoh@ornl.gov}\affiliation{Neutron Scattering Division, Oak Ridge National Laboratory, Oak Ridge, Tennessee 37831, USA }

%\date{}

\begin{abstract}
Through analysis of single crystal neutron diffraction data, we present the magnetic structures of magnetoelectric Co$_4$Nb$_2$O$_9$ under various magnetic fields. In zero-field, neutron diffraction experiments below $T_N$=27 K reveal that the Co$^{2+}$ moments order primarily along the $a^*$ direction without any spin canting along  the $c$ axis, manifested by the magnetic symmetry $C2/c'$. The moments of nearest neighbor Co atoms order ferromagnetically with a small cant away from the next nearest neighbor Co moments along the $c$ axis. In the applied magnetic field H$\parallelsum$ $a$, three magnetic domains were aligned with their major magnetic moments perpendicular to the magnetic field with no indication of magnetic phase transitions. The influences of magnetic fields on the magnetic structures associated with the observed magnetoelectric coupling are discussed.
\end{abstract}


\maketitle
\section{INTRODUCTION}

The cross coupling of magnetization and electric polarization to their conjugate magnetic and electric field, well-known as magnetoelectric (ME) effect\citep{schmid1994, rivera2009}, has drawn a great deal of interest due to not only its essential role in the quest for emergent states of matter\citep{tokura2006, essin2009} and novel types of ferroic order \citep{spaldin2008, schmid2008, fiebig2005, tokura2014} in condensed matter physics but also potential applications in spintronics\citep{wang2003, spaldin2005, cheong2007, eerenstein2006, tokura2014}.  
The microscopic mechanisms underpinning the ME effect remain $hitherto$ unsettled but basically they are restricted by symmetry \citep{schmid1994, fiebig2005}. Group theory requires that specific symmetry elements, namely spatial inversion and time reversal, have to be broken to make the ME effect active. 
Such symmetry restrictions are also applicable to a novel ferrotoroidic order that is related to the antisymmetric part of the linear ME tensor \citep{spaldin2008, schmid2008}. Materials showing ferrotoroidicity are elusive and a few typically proposed and investigated systems are the metal orthophosphates such as LiCoPO$_4$ \citep{van2007}, MnPS$_3$ \citep{ressouche2010}, and the pyroxenes such as CaMnGe$_2$O$_6$ \citep{ding2016a} and LiFeSi$_2$O$_6$ \citep{baum2013}.

\begin{figure}
\centering
\includegraphics[width=1\linewidth]{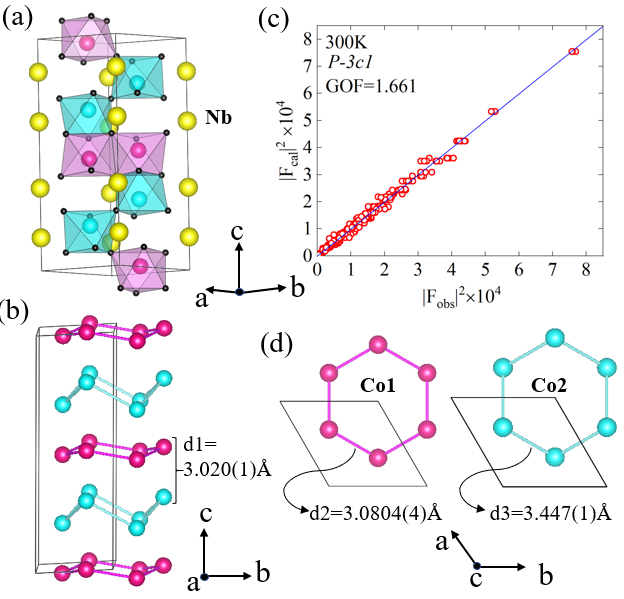}
\caption{(Color Online) (a) The crystal structure of Co$_4$Nb$_2$O$_9$ at room temperature. Co1 and Co2 atoms (pink and cyan spheres, respectively) are shown within their octahedral oxygen (black spheres) coordination. Nb atoms are represented by yellow large spheres. (b) and (d) highlight the buckled Co1 and Co2 honeycomb layers that are stacked along the $c$ axis with the marked essential bonding distances between Co atoms. (c) Results of the crystal structure refinement from the x-ray data at room temperature.}\label{fig:1}
\end{figure}

Recently a corundum-type compound, Co$_4$Nb$_2$O$_9$, has been found to show both large ME and magnetodielectric effect below the N\'eel temperature $T_N\approx$  27 K \cite{Fischer1972, kolo2011, fang2014, khanh2016, yin2016}. More interestingly, both the electric-field induced magnetization and magnetic field controlled polarization have been experimentally observed on a powder sample\cite{kolo2011, fang2014}. Co$_4$Nb$_2$O$_9$ crystallizes with the $\alpha$-Al$_2$O$_3$-type trigonal crystal structure with the space group $P\bar{3}c1$ \citep{bertaut1961}(see Fig.\ref{fig:1}a) and can be viewed as a derivative of Cr$_2$O$_3$, one of the first predicted and discovered, and intensively studied room temperature ME materials \cite{Dzyaloshinskii1959, astrov1960, mcgurie1956, fiebig1994, kimura2013}. The magnetic structure of Co$_4$Nb$_2$O$_9$ was first determined by Bertaut $et al.$\cite{bertaut1961, schwarz2010} to have antiferromagnetically coupled ferromagnetic Co$^{2+}$ chains with the moments along the $c$-axis. The determined magnetic symmetry allows a linear ME effect but is incompatible with the magnetoelectric effect recently measured on a single crystal \citep{khanh2016}. Recently, a different magnetic structure, in which all spins are nearly parallel to the [1$\bar{1}$0] direction with a canting along the $c$ axis, was suggested based on a single crystal neutron diffraction experiment\citep{khanh2016}. Later on, Deng et al. argued another distinct magnetic structure without any spin canting to the $c$ axis from powder neutron diffraction data \citep{deng2018}. 

Moreover, different spin-flop behaviors have been observed in Co$_4$Nb$_2$O$_9$ at a relatively small magnetic field of 0.2 T along the [1$\bar{1}$0] direction in a single crystal sample \citep{khanh2016} and 1.2 T in a powder sample \citep{fang2014, kolo2011}. Then, conjectured magnetic structures associated with these magnetic anomalies have been suggested to explain the large ME and magnetodielectric effect \citep{khanh2016, kolo2011}. More interestingly, Khanh et al. have found that the electric polarization vector can be promptly controlled by applying an in-plane magnetic field\citep{khanh2017}. They attributed this effect to the continuous rotation of the antiferromagnetic moments on the honeycomb lattice.  

To understand the large ME effect, one has to know the precise magnetic structure especially in such a complex systems where magnetic properties are dominated by both the exchange interactions and single-ion anisotropy \citep{deng2018}. The magnetic structures in magnetic fields, which may explain the robust manipulation of the electric polarization by a magnetic field, remain unknown. In light of this, we set out to revisit the magnetic and magnetoelectric effect of Co$_4$Nb$_2$O$_9$ in different magnetic fields on a well-characterized high quality single crystal.  

In this work, we report the detailed magnetic and magnetoelectric properties measured on a high quality Co$_4$Nb$_2$O$_9$ single crystal and the evolution of magnetic structures with temperature and magnetic field measured by single crystal neutron diffraction. We show that at zero magnetic field, single crystal neutron diffraction data unveil a magnetic structure with the ordered moments only confined in the $ab$ plane, which allows a linear ME effect as further confirmed by the electric polarization measurements. We discuss the influence of magnetic field on the ME effect with neutron diffraction data under various magnetic fields.

\begin{table*} 
\caption{The structure parameters of Co$_4$Nb$_2$O$_9$ measured at 260 K by single crystal x-ray diffraction. The space group is $P\bar{3}c1$, $a$=5.1877(4) \AA,  $b$=5.1877(4) \AA, $c$=14.1841(4) \AA, $\alpha$=90$^o$, $\beta$=90$^o$, $\gamma$=120$^o$. $R_f$=3.2\%. $\chi^2$=1.66.   $U$ in \AA$^2$.}\label{str}
\begin{tabular}{cccccccccccc}
\hline
\hline
atom & $type$ &$site$& $x$& $y$ &  $z$  & $U_{11}$ & $U_{22}$ & $U_{33}$ & $U_{12}$ & $U_{13}$ & $U_{23}$\\
\hline
Nb1 & Nb  &  $4c$ &  0& 0&  0.3578(1) & 0.0079(8) & 0.0079(8) & 0.004(1) & $U_{11}$/2 & 0 & 0 \\
Co1 & Co  &  $4d$ &  1/3 & 2/3&  0.0138(3) & 0.007(2) & 0.007(2) & 0.0012(3) & $U_{11}$/2 & 0 & 0 \\
Co2 & Co  &  $4d$ &  1/3 & 2/3&  0.3071(2) & 0.007(2) & 0.007(2) & 0.013(2) & $U_{11}$/2 & 0 & 0 \\
O1  & O   &  $6f$ &  0.297(3) & 0&  1/4 & 0.015(8) & 0.004(8) & 0.007(6) & $U_{22}$/2 & -0.001(5) & -0.001(5) \\
O2  & O   & $12g$ &  0.342(1) & 0.309(3)& 0.0838(7) & 0.011(6) & -0.012(4) & 0.014(4) & 0.005(4) & 0.007(5) & 0.005(4) \\
\hline
\hline
\end{tabular}
\end{table*}

\begin{table} 
\caption{The structure parameters of Co$_4$Nb$_2$O$_9$ measured at 50 K by single crystal neutron diffraction. The space group is $P\bar{3}c1$, $a$=5.180(3) \AA,  $b$=5.180(3) \AA, $c$=14.163(6) \AA, $\alpha$=90$^o$, $\beta$=90$^o$, $\gamma$=120$^o$. $R_f$=4.51\%. $\chi^2$=3.64. The atomic displacement parameter $B_{iso}$ is in 1/(8$\pi^2$)\AA$^2$.}\label{strND}
\begin{tabular}{ccccccc}
\hline
\hline
atom & $type$ &$site$& $x$& $y$ &  $z$  & $B_{iso}$ \\
\hline
Nb1 & Nb  &  $4c$ &  0& 0&  0.3572(6) & 0.4 \\
Co1 & Co  &  $4d$ &  1/3 & 2/3&  0.018(1) & 0.4 \\
Co2 & Co  &  $4d$ &  1/3 & 2/3&  0.3068(9) & 0.4\\
O1  & O   &  $6f$ &  0.296(3) & 0&  1/4 & 0.6\\
O2  & O   & $12g$ &  0.346(2) & 0.303(3)&  0.0842(8) & 0.6\\
\hline
\hline
\end{tabular}
\end{table}

\section{EXPERIMENTAL METHODS}
Single crystals of Co$_4$Nb$_2$O$_9$ were grown by the traveling-solvent floating-zone (TSFZ) technique. The feed and seed rods for the crystal growth were prepared by solid state reaction. Appropriate mixtures of CoCO$_3$, and Nb$_2$O$_5$ were ground together and pressed into 6 mm diameter $\times$ 60 mm rods under 400 atm hydrostatic pressure and then calcined in air at 1000 atm for 24 h. The crystal growth was carried out in argon in an IR-heated image furnace (NEC) equipped with two halogen lamps and double ellipsoidal mirrors with feed and seed rods rotating in opposite directions at 25 rpm during crystal growth at a rate of 4 mm/h.

Single-crystal x-ray diffraction data were collected at 260 K using a Rigaku XtaLAB PRO diffractometer with the graphite monochromated Mo $K_{\alpha}$ radiation ($\lambda$ = 0.71073 \AA) equipped with a Dectris Pilatus 200 K detector and an Oxford N-HeliX cryocooler. Peak indexing and integration were done using the Rigaku Oxford Diffraction CrysAlisPro software \citep{rigaku}. An empirical absorption correction was applied using the SCALE3 ABSPACK algorithm as implemented in CrysAlisPro \citep{higashi2000}. Structure refinement was done using FullProf Suite \cite{fullprof}.

The dc magnetization curves were obtained using a high-field vibrating sample magnetometer (VSM) of the National High Magnetic Field Laboratory. For the ac susceptibility measurement, the conventional mutual inductance technique was used with frequencies below 1000 Hz. Two balanced sensing coils were prepared and the sample was inserted into one of the two sensing coils. When the sample was magnetized by small ac magnetic field superimposed on external dc magnetic field, the magnetic susceptibility signal appears as unbalanced voltages across the sensing coils, which were measured using lock-in amplifiers. 

For the dielectric constant and the electric polarization measurements, single crystal samples were used and the orientations were determined by Laue diffraction. Two single crystalline samples were polished to achieve two parallel flat surfaces perpendicular to the $a$-axis and the $c$-axis. The dimensions were $1.8 \times 1.8 \times 0.3$ mm$^3$ and $2.2 \times 2.2 \times 0.4$ mm$^3$, respectively. An Andeen-Hagerling AH-2700A commercial capacitance bridge was used to measure the capacitance, which was converted to dielectric constant using the relation between the capacitance and an infinite parallel capacitor. The electric polarization was obtained by integrating the pyroelectric current ($I_p$) with respect to time. The $I_p$ was measured during warm up after the sample was cooled in the presence of the poling electric field and/or external magnetic field. The detailed procedure can be found in Ref.\cite{lee2014}.

Single-crystal neutron diffraction was performed at the HB-3A Four-Circle Diffractometer (FCD) equipped with a 2D detector at the High Flux Isotope Reactor(HFIR) at Oak Ridge National Laboratory (ORNL).
A neutron wavelength of 1.003~\AA~ (neutron energy 81 meV) was used with a bent perfect Si-331 monochromator \cite{hb3a}. The nuclear and magnetic structure refinements were performed with the FullProf Suite \cite{fullprof}.
Single crystal neutron diffraction under various magnetic fields was measured at a cold neutron triple axis spectrometer (CTAX) at HFIR at ORNL. The neutron wavelength of 4.045 \AA~(neutron energy 5 meV) was used.

\section{RESULTS}
\subsection{CRYSTAL STRUCTURE}
In view of the inconclusive magnetic structure and the debatable spin-flop behavior of Co$_4$Nb$_2$O$_9$, it is important to examine the as-grown crystals as effects such as crystal domains, impurities or defects can often give rise to unpredictable and spurious results. Hence we first carefully characterized our single crystals using x-ray single crystal diffraction. The crystal structure was solved and refined based on the x-ray single crystal diffraction data at 260 K as described in the experimental section. More than 2290 reflections (effective reflections 560 with I $>$ 4* $\sigma$) were used in the structure refinements. It shows that Co$_4$Nb$_2$O$_9$ crystallizes with the space group $P\bar{3}c1$, in good agreement with the previous report \citep{bertaut1961}. The single crystal was practically perfect without suffering from twins, impurities and significant defects. The data fit quality is shown in Fig. \ref{fig:1}(c) with a comparison between the observed squared structure factor and the calculated one with a goodness-of-fit value of 1.66. 

We further measured the Co$_4$Nb$_2$O$_9$ single crystal using single crystal neutron diffraction at 50 K which confirms the space group $P\bar{3}c1$ and reveals a good quality of the single crystal. The structural parameters from the refinement of single crystal x-ray and neutron diffraction data were summarized into Table \ref{str} and \ref{strND}, respectively. As presented in Fig. \ref{fig:1}, the crystal structure consists of an alternate stacking of slightly buckled Co1 honeycomb and  buckled Co2 honeycomb layers along the $c$-axis. Even though Co$_4$Nb$_2$O$_9$ features a stacking of honeycomb layers, the nearest-neighbor bonding between Co cations is the Co1-Co2 bond (d1=3.020(1)\AA) along the $c$-axis as marked in Fig. \ref{fig:1}(b). The nearest bonding between Co atoms within each honeycomb layer is d2=3.0804(4)\AA~and d3=3.447(1)\AA~for for Co1-Co1 and Co2-Co2 bonds, respectively(Fig. \ref{fig:1}(d)).   

\subsection{MAGNETIC AND MAGNETOELECTRIC PROPERTIES}
\begin{figure}
\centering
\includegraphics[width=1\linewidth]{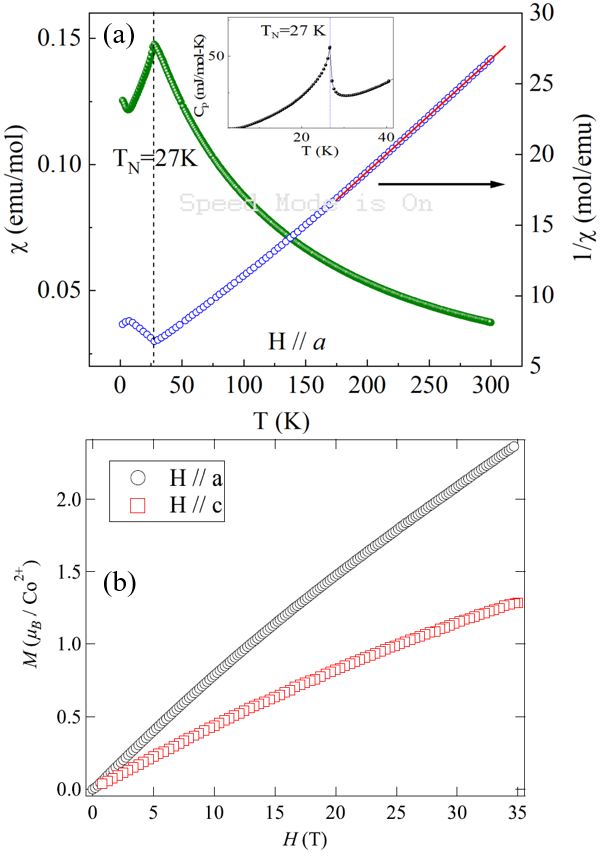}
\caption{(Color Online) (a) The magnetic susceptibility of Co$_4$Nb$_2$O$_9$ as a function of temperature with H =0.1 T parallel to the $a$-axis at 1.7 K. Inset shows the temperature-dependent heat capacity of Co$_4$Nb$_2$O$_9$. (b) The isothermal magnetization curves at 1.5 K with magnetic field up to 35 T.}\label{fig:2}
\end{figure}

Magnetic susceptibility measurement shows that Co$_4$Nb$_2$O$_9$ undergoes an antiferromagnetic transition at $T_N$=27 K, characterized by a sharp cusp in the susceptibility curve(Fig. \ref{fig:2}), in accordance with the previous results \citep{khanh2016}. The Curie-Weiss fit of the inverse susceptibility data from 175 K to 300 K yields an effective moment 5.0$\mu_B$ that indicates a high spin-state for Co$^{2+}$ with $S$ = 3/2. To characterize the magnetic phase transition, we have for the first time measured the heat capacity of Co$_4$Nb$_2$O$_9$. It reveals a sharp peak in the heat capacity curve, reflecting the nature of long range spin order (Fig. \ref{fig:2}(a)). 

\begin{figure*}
\centering
\includegraphics[width=0.8\linewidth]{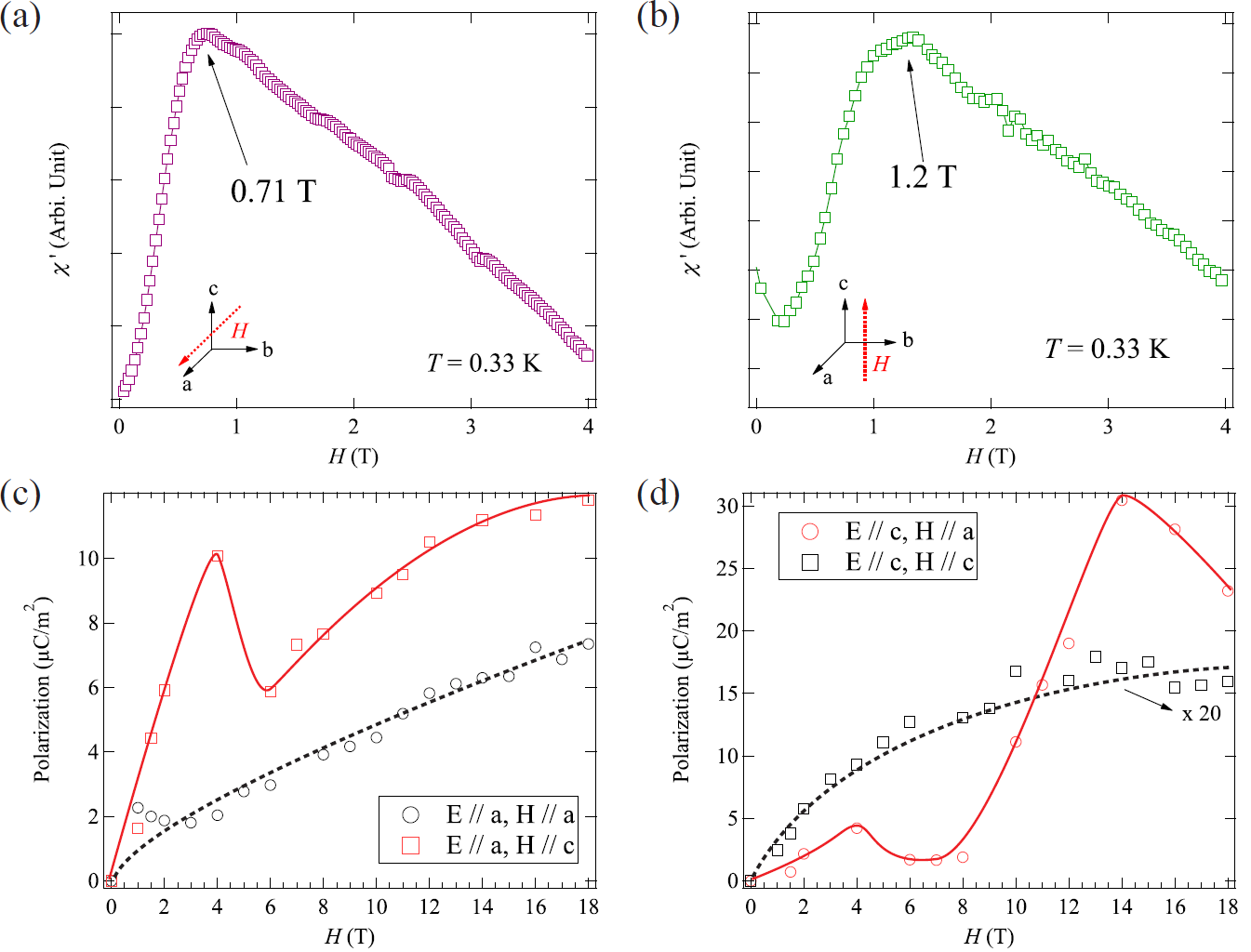}
\caption{(Color Online) (a) and (b) The magnetic-field-dependent ac susceptibility with H parallel to the $a$- and $c$ axes at 0.33 K, respectively. (c) and (d) The magnetic field dependence of electric polarization at 21.5 K with different magnetoelectric annealing configurations. The lines are guides to the eye.}\label{fig:3}
\end{figure*}
The isothermal magnetization measurements at high magnetic fields were carried out at 1.5 K up to 35~T along both the $a$ axis and the $c$ axis. No magnetic saturation was observed. The magnetization along the $a$-axis is evidently higher than that along the $c$-axis suggesting an easy-plane magnetic anisotropy \citep{khanh2016}. Up to 10 T, the magnetization along both the $a$- and $c$ axes show a near linear behavior.  
The ac magnetic susceptibilities as a function of magnetic field were measured at various temperatures with the applied magnetic field parallel to the $a$ and $c$ axes. As shown in Fig. \ref{fig:3}(a) and (b), the ac susceptibility curves at T=0.33 K reveal two critical magnetic fields of 0.71~T and 1.2~T along the $a$- and $c$-axes, respectively. Such anomalies were tentatively explained as ``spin flop" behavior in the earlier reports \citep{kolo2011, fang2014,solovyev2016}. However, as we concluded by our field-dependent neutron diffraction data later, these features are attributed to a combined consequence of alignment of magnetic domains and spin rotations. 

To detect the magnetoelectric effect, the electric polarization of Co$_4$Nb$_2$O$_9$ single crystal was measured under various magnetic and electric fields along the $a$- or $c$- axes after completing the magnetoelectric annealing procedure \citep{ding2016a}. The detailed field-induced pyroelectric current and electric polarization curves as a function of temperature at various magnetic fields are shown in Fig. S1 and S2. Figure \ref{fig:3}(c) and (d) show the magnetic-field-dependent electric polarization measured at 21.5 K using electric fields E=320 kV/m and 260 kV/m, respectively.  When the applied magnetic field is perpendicular to the electric field, the value of polarization increases linearly below 4 T with the increase of magnetic field. However, a steep drop sets in around the applied magnetic field 4 T. This is a phenomenon that has not been found in the previous works, likely associated with the procedure of the alignment of magnetic domains as explained in the following. Above this magnetic field, the electric polarization restores the linear behavior. It is clear that when both the magnetic and electric field are applied in the same direction, either the $a$ or $c$ axis, relatively smaller polarization is observed. This indicates that the off-diagonal ME tensor components have larger values than the diagonal ones, allowed by the magnetic symmetry $C2/c'$ (see neutron diffraction section). Evidently, the magnetoelectric properties in Co$_4$Nb$_2$O$_9$ are distinct from that in Cr$_2$O$_3$ where only the diagonal ME tensor terms were observed below the spin-flop field imposed by its magnetic symmetry $P\bar{3}'m'$\cite{astrov1960,shirane1965}.  Therefore, such an observation reflects that the previously documented magnetic structure with the magnetic point group $-3'm'$ which allows only diagonal terms is inappropriate \cite{bertaut1961, schwarz2010}. 

\begin{figure*}
\centering
\includegraphics[width=1\linewidth]{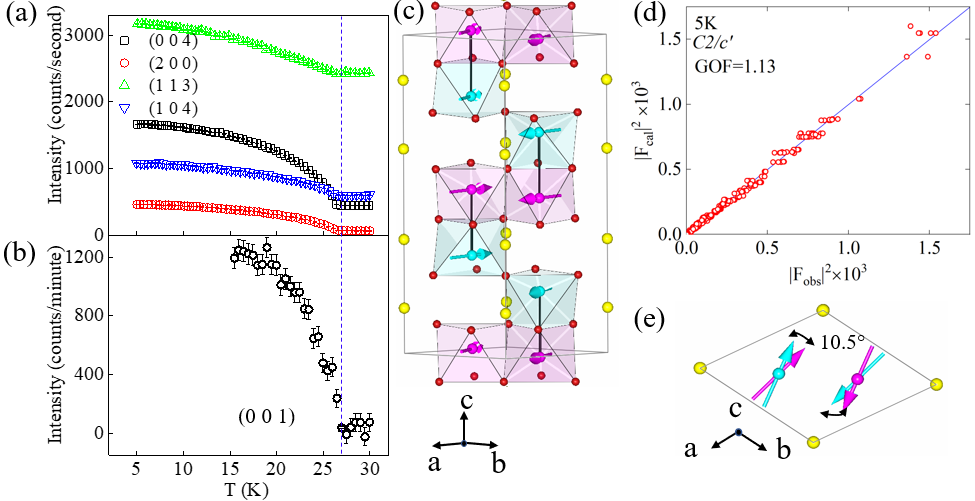}
\caption{(Color Online) (a-b) The representative magnetic reflections as a function of temperature reflecting the magnetic phase transition at 27 K. The data were measured at HB-3A with the neutron wavelength of 1.003 \AA. (c) The magnetic structure of Co$_4$Nb$_2$O$_9$ showing the ferromagnetically coupled Co1-Co2 pairs (the nearest neighbor Co1 and Co2 atoms). (d) Results of the magnetic structure refinement for the neutron data collected at 5 K. (e) All spins are confined in the $ab$ plane with the canting angle between Co1 and Co2 spins (the next-nearest-neighbor along the $c$ axis).}\label{fig:4}
\end{figure*}

\subsection{ZERO-MAGNETIC-FIELD SINGLE CRYSTAL NEUTRON DIFFRACTION}
We performed single crystal neutron diffraction to revisit the magnetic structure of Co$_4$Nb$_2$O$_9$ down to 5 K. As shown in Fig. \ref{fig:4}, several selected Bragg reflections were measured to track the magnetic phase transition upon warming. One can clearly see the increased intensity for the representative reflections (0 0 4), (2 0 0), (1 1 3), (1 0 4)   and (0 0 1) below $T_N$. This indicates that all the magnetic reflections can be well indexed by a propagation vector \textbf{k}=\textbf{0}, as reported in Ref. \citep{khanh2016, deng2018}. No further crystal structure transition was detected down to 5 K. 

Symmetry analysis was performed to determine the magnetic structures of Co$_4$Nb$_2$O$_9$. A number of magnetic subgroups that are compatible with the given space group and propagation vector can be calculated by Bilbao Crystallographic Server (Magnetic Symmetry and Applications \citep{bilbao}) software. The magnetic structures bearing trigonal symmetry imply the magnetic moments along the $c$-axis, inconsistent with our experimentally measured magnetic reflections such as (0 0 4) and (0 0 1). Therefore, we have to lower the symmetry from the k-maximal magnetic symmetry in the subgroup hierarchy. We then found the monoclinic magnetic subgroups $C2'/c'$, $C2/c'$, $C2'/c$ and $C2/c$. The observed magnetic reflections also imply that spins should be confined into the $ab$ plane, distinct from the previous report where a magnetic moment component out of the $ab$ plane was found \citep{khanh2016}. 
Since the magnetic peaks superpose on the nuclear reflections below $T_N$, combined magnetic and nuclear structure refinement was performed. The refinement using $C2/c'$ magnetic space group yields the best fitting with $R_f$=6.4\% and $\chi^2$=1.13 for the combined phase for 745 reflections. The results of the refinement and the corresponding illustrations of magnetic structure are shown in Fig.\ref{fig:4}. The magnetic structure solved can be simply viewed as antiferromagnetically coupled ferromagnetic chains along the $c$-axis. Magnetic moments in each chain are confined into the $ab$ plane with ferromagnetic pairs for the nearest neighbor Co1 and Co2 atoms and a canting angle 10.5$^{\circ}$ between the next-nearest-neighboring Co1-Co2 spins in the chain. The ordered magnetic moment is 2.820(8) $\mu_B$ (m$_x$=3.201(8)$\mu_B$, m$_y$=2.12(1)$\mu_B$, m$_z$=0) for both the Co1 and Co2 sites. The magnitude of the ordered magnetic moment for Co$^{2+}$ is close to the theoretical spin-only ordered value 3 $\mu_B$ for Co$^{2+}$ with a high spin state, indicating that the orbital moment is nearly quenched as shown in \citep{ding2016b, hutanu2012}. Note that due to the existence of the trigonal lattice symmetry, three magnetic domains were considered and their populations were set to be equal during the refinement. In the following, we show that the population of magnetic domains in Co$_4$Nb$_2$O$_9$ can be tuned by applying an external magnetic field. 

The magnetic symmetry can also be readily derived by the observed magnetoelectric coupling. By applying the Neumann’s principle to the magnetoelectric effect with these possible magnetic symmetries, we could exclude the subgroups $C2'/c'$ and $C2/c$. This is because, in principle, neither of them allows the presence of the linear ME effect. The $C2'/c'$ and $C2/c$ subgroups do not necessarily mean a ferromagnetic structure (in other words, there is no symmetry constraint for a ferromagnetic arrangement); hence this is not the reasoning that we ruled them out \citep{khanh2016}. The observed ME properties can help us preclude the $C2'/c$ subgroup since this symmetry allows only the off-diagonal terms. Thereby, the only magnetic subgroup compatible with Co$_4$Nb$_2$O$_9$ is $C2/c'$ based on our symmetry analysis and experimental observations, in good agreement with the neutron diffraction results. 

\begin{figure}
\centering
\includegraphics[width=1\linewidth]{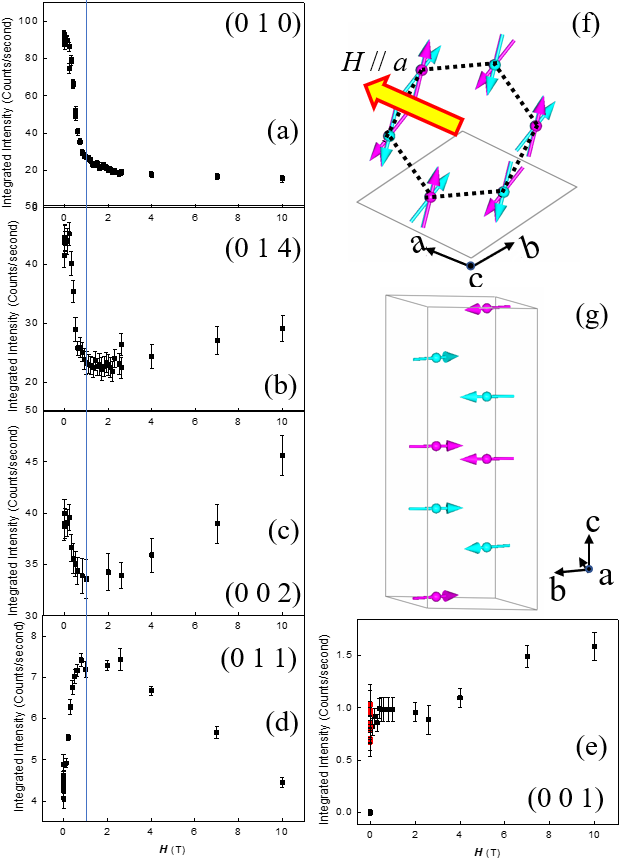}
\caption{(Color Online) (a)-(e) The selected magnetic reflections as a function of magnetic field along the $a$-axis measured at CTAX. The solid line marks the alignment of magnetic domains. (f-g) The magnetic structure of Co$_4$Nb$_2$O$_9$ after the three magnetic domains are aligned. All the Co spins are nearly parallel to the $b^*$ direction.}\label{fig:5}
\end{figure}

\subsection{SINGLE CRYSTAL NEUTRON DIFFRACTION IN MAGNETIC FIELDS}
To clarify the hidden mechanism of the complex ME effect in Co$_4$Nb$_2$O$_9$, we carried out neutron diffraction experiments under high magnetic fields at CTAX. The same crystal used at HB-3A was measured in the (0 K L) scattering plane. The crystal misalignment was less than 1$^{\circ}$ for this experiment. The magnetic field was applied vertically, i.e., along the crystal's $a$-axis. The representative reflections were measured with rocking curve scans at selected temperatures under magnetic field up to 10~T. The integrated magnetic intensities of these reflections are plotted versus the magnetic field in Fig. \ref{fig:5} (a-e). They were obtained by subtracting the integrated intensities at 50 K from that at 1.5~K. We found a considerable decrease in the intensities of magnetic reflections (0 1 0) and (0 1 4) with the increasing magnetic field up to 1 T and a significant increase in the intensity of magnetic reflection (0 1 1). There is only a slight decrease in intensity of the magnetic peak (0 0 2). These changes indicate a spin reorientation from the trigonal lattice direction [1 0 0], [0 1 0] or [1 1 0] to the direction parallel the reciprocal [0 1 0] direction (the $b^*$-axis), which causes a large decrease of the magnetic scattering at (0 1 0), (0 1 4) reflections but an evident increment of the (0 1 1) magnetic reflection. The observed weak intensity at 1~T and above can be related to the induced moments along the field direction and also the slight misalignment of the crystal. The magnetic intensities at $H$=1 T were obtained by scaling them to those measured at zero magnetic field that was well solved with the complete data measured at HB-3A for catching all the necessary corrections. With these magnetic reflections at 1~T, the magnetic structure with a single magnetic domain can fit the data in a satisfactory quality ($\chi^2$=6.09) and yields the moments along [1 2 0] direction in real space and  [0 1 0] direction in reciprocal space (Fig.\ref{fig:5}), in the same magnetic symmetry $C2/c'$  as that at zero magnetic field. The ordered magnetic moments are 2.6(1) $\mu_B$ for both Co sites with a slightly increased planar canting angle of 12.2$^{\circ}$. The magnetic field of 1 T along the $a$-axis does not significantly change the moment size and the magnetic symmetry but switches three magnetic domains with spins along the three crystal axes to the single one with the spins along the reciprocal $b^*$-axis and increases the canting angle slightly.
The above refinement did not consider the induced ferromagnetic moments along the $a$-axis (the induced moments are small as expected from the magnetization measurements in Fig. \ref{fig:2}).  

Magnetic field 1 T$<$H$<$4 T does not greatly change the intensities of magnetic reflections (0 1 0), (0 1 4), (0 0 2), (0 1 1) and (0 0 1), suggesting a rather robust magnetic ground state. Above 4 T, only the magnetic reflection (0 1 1) was appropriately suppressed, reflecting no further magnetic phase transition but spin rotation in the $ab$ plane. Indeed, by refining the observed magnetic reflections at 10 T using the magnetic symmetry obtained at 1 T, we arrived at a smaller planar canting angle 8.5$^{\circ}$ and magnetic moment 2.9(3)$\mu_B$ for both sites. In the magnetic-field-dependent polarization curves, it is clear that a steep drop occurs around 4 T, coincident with the abrupt decrease of the observed magnetic reflection (0 1 1). Above 4 T, the restored linear magnetoelectric effect supports our conclusion that a magnetic field up to 10 T does not break the magnetic symmetry but rotates the magnetic moments in the $ab$ plane. 

\section{DISCUSSION}
Our work reveals that Co$_4$Nb$_2$O$_9$ antiferromagnetically orders below 27 K with the magnetic space group $C2/c'$ which allows a linear ME effect. This point is in good agreement with the previous reports by Khanh et al. \citep{khanh2016} and Deng et al.\citep{deng2018}. However, in the former case, they found that all spins are exactly ferromagnetically aligned with a considerable canting angle 22$^{\circ}$ toward the $c$-axis(mainly in the $ab$ plane).  This in fact conflicts with our results in which spin out of the $ab$ plane is minimal based on our neutron diffraction data even though m$_z$ component is symmetry allowed. In the latter case, a distinct magnetic structure was reported where all moments are purely in the $ab$ plane with the canting angle of 1.3$^{\circ}$ and 25.2$^{\circ}$ for the neighbor Co1 and Co2 moments, respectively. 
By contrast, we found a magnetic structure akin to the latter case but with a different configuration by a careful neutron diffraction measurement on a high quality single crystal. As illustrated in Fig. \ref{fig:4}, the nearest-neighbor Co1 and Co2 atoms form a ferromagnetic pair in the $ab$ plane and the essential (0 0 1) magnetic reflection ensues a planar canting angle between each adjacent Co1-Co2 pair. This makes the canting angle for the neighbor Co1 and Co2 moments equivalent.
Our magnetic arrangement seems to be more favored by the special crystal structural arrangement of Co atoms since the nearest-neighbor bonding between Co atoms is the edge-sharing Co1O$_6$ and Co2O$_6$ octahedra with a bond distance of d1=3.020(1)\AA ~along the $c$ axis as marked in Fig. \ref{fig:1}(b). The dominant ferromagnetic exchange interactions of Co1 and Co2 (d1) comparing to other exchange strength promotes a canting angle between the magnetic moments of Co1 or Co2 neighbors \citep{deng2018}. Moreover, the strong single-ion anisotropy of Co in this system should also play a vital role in forming such a magnetic configuration.
 
When applying a magnetic field H$<$1 T, the magnetic symmetry is practically invariant but with an alignment of the three magnetic domains. The magnetic moments are nearly parallel to the $b^*$ direction, elucidating the conjectured ''spin flop" suggestions in the previous works\citep{khanh2016, fang2014, kolo2011}. In the magnetic field dependence of electric polarization curves, a sudden drop around 4 T is likely a consequence of the completion of the alignment of the magnetic domains since thermally stimulated current sources in the course of magnetic domain alignment may contribute to the pyroelectric current so as to the electric polarization\citep{zou2014, ngo2015}. Above 4 T, i.e., in a single magnetic domain, the observed magnetoelectric effect seems to be more intrinsic because we have confirmed that the magnetic structure up to 10 T remains the same symmetry with a spin rotation within the $ab$ plane. Such a robust magnetic symmetry in a magnetic field, that probably emanates from the strong easy-plane single-ion anisotropy, provides a natural explanation of the manipulation of electric polarization \citep{khanh2017, solovyev2016}.

The solved magnetic symmetry with its off-diagonal components non-null in Co$_4$Nb$_2$O$_9$ in principle allows the occurrence of the ferrotoroidal order. The experimental observation of the relevant ferrotoroidal domains through a magnetoelectric annealing process is greatly desired. This can be spatially resolved by optical second harmonic generation and spherical neutron polarimetry determination of the relative domains populations \citep{van2007, ressouche2010}.

\section{CONCLUSION}
We have investigated the magnetoelectric properties and magnetic structure evolution with temperature and magnetic field of a high quality single crystal Co$_4$Nb$_2$O$_9$. Single crystal neutron diffraction without magnetic field revealed a magnetic structure below 27 K characterized by the $C2/c'$ symmetry with magnetic moments totally confined in the $ab$ plane, which allows the linear ME effect as observed experimentally by the pyroelectric current measurements. Single crystal neutron diffraction in magnetic field with H $\parallelsum$ a showed that the magnetization anomaly around 1 T is in fact a magnetic domain alignment that drives the magnetic moments parallel to the $b^*$ direction without breaking the magnetic symmetry. A higher magnetic field up to 10 T did not change the magnetic symmetry but rotates the magnetic moments in the $ab$ plane. The robust magnetic symmetry to the external magnetic field offers a natural way to manipulate and control the electric polarization in this system.

\begin{acknowledgments}
LD thanks J. Rodriguez-Carvajal for helpful discussions. The research at Oak Ridge National Laboratory (ORNL) was supported by the U.S. Department of Energy (DOE), Office of Science, Office of Basic Energy Sciences, Early Career Research Program Award KC0402010, under Contract DE-AC05-00OR22725 and the U.S. DOE, Office of Science User Facility operated by the ORNL.
The work at University of Tennessee was supported by DOE under award DE-SC-0020254. A portion of this work was performed at the National High Magnetic Field Laboratory, supported by the National Science Foundation Cooperative Agreement No. DMR-1644779 and the State of Florida. 
The US Government retains, and the publisher, by accepting the article for publication, acknowledges that the US Government retains a nonexclusive,
paid-up, irrevocable, worldwide license to publish or reproduce
the published form of this manuscript, or allow others to do so, for US Government purposes. The Department of Energy will provide public access to these results of federally sponsored research in accordance with the DOE Public Access
Plan.\citep{DOE}.
\end{acknowledgments}

\clearpage

\section{Appendix: Polarization measurements of Co$_4$Nb$_2$O$_9$ }

\begin{figure*}
\linespread{1}
\par
\renewcommand{\thefigure}{S1}
\begin{center}
\includegraphics[width=5in]{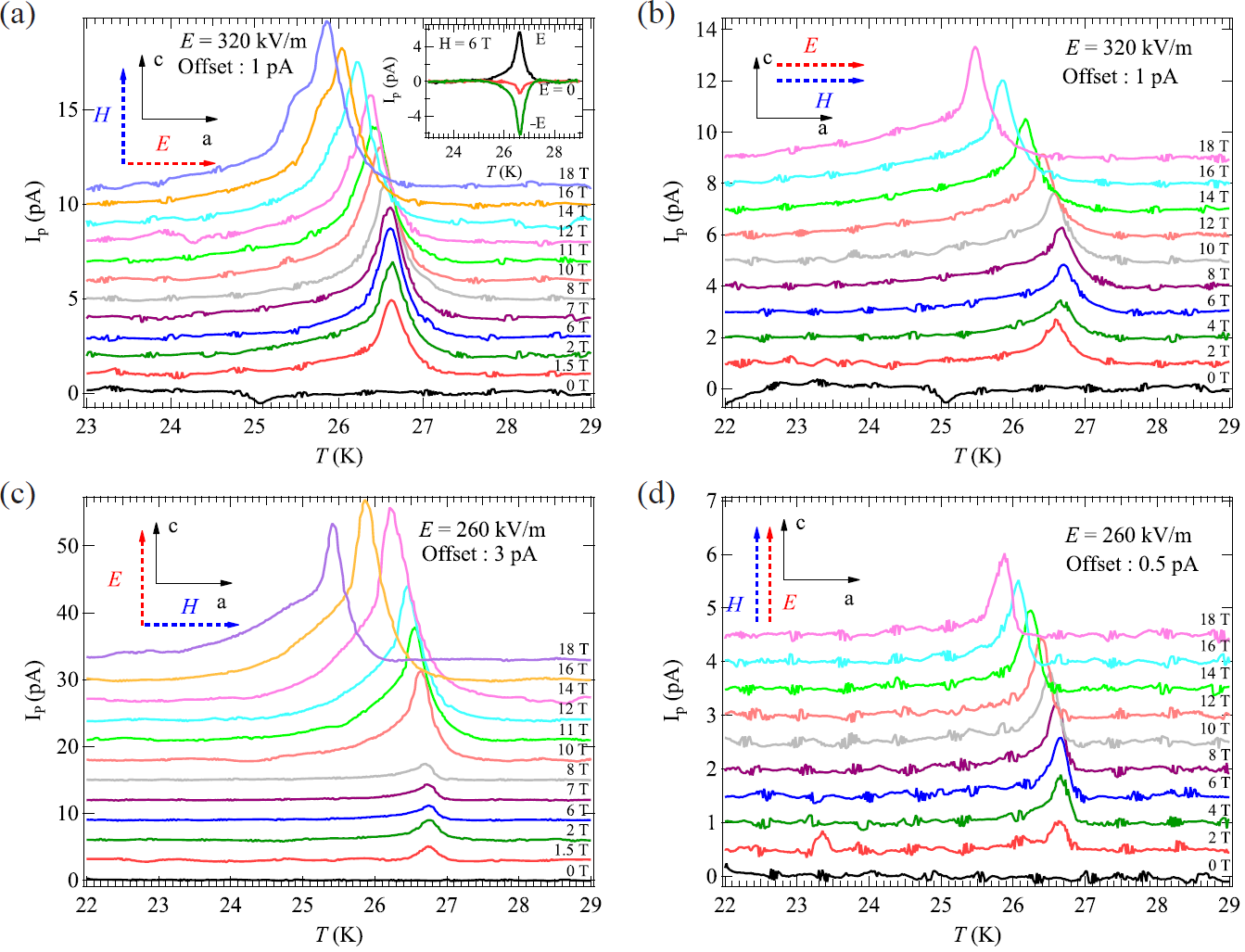}
\end{center}
\par
\caption{(Color Online) Pyroelectric current ($I_p$) measured under poling field $E$ and magnetic field $H$. Directions of the applied  $E$ and $H$ are shown by arrows along either the $a$- or $c$-axis.}
\end{figure*}

\begin{figure*}
\linespread{1}
\par
\renewcommand{\thefigure}{S2}
\begin{center}
\includegraphics[width=5in]{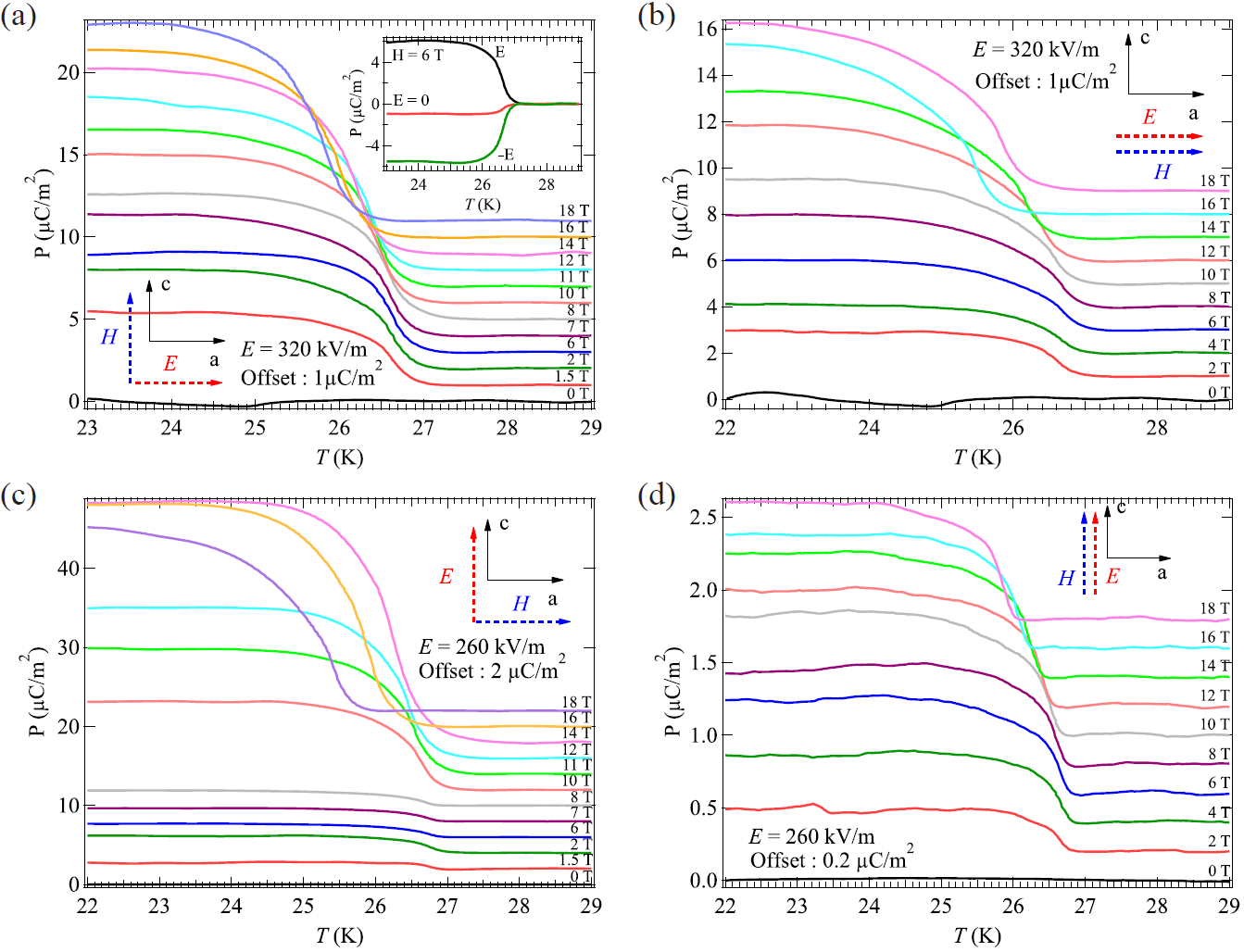}
\end{center}
\par
\caption{(Color Online) Polarization data are obtained by integrating $I_p$ shown in Fig.6. Directions of the poling field $E$ and magnetic field $H$ are shown by arrows along either the $a$- or $c$-axis.}
\end{figure*}


\begin{thebibliography}{99}

\bibitem{schmid1994} H. Schmid, Ferroelectrics 162, 317 (1994).
\bibitem{rivera2009} J.-P. Rivera, Eur. Phys. J. B 71, 299–313 (2009)  % A short review of the magnetoelectric effect and related experimental techniques on single phase (multi-) ferroics
\bibitem{tokura2006} Y. Tokura, Science, 312, 1481 (2006).
\bibitem{essin2009} A. M. Essin, J. E. Moore, and D. Vanderbilt, Phys. Rev. Lett. 102, 146805 (2009).
\bibitem{fiebig2005} M. Fiebig, J. Phys. D 38, R123 (2005). % Revival of the magnetoelectric effect
\bibitem{tokura2014} Y. Tokura, S. Seki, and N. Nagaosa, Rep. Prog. Phys. 77, 076501 (2014).
\bibitem{spaldin2008} N. A. Spaldin, M. Fiebig, and M. Mostovoy, J. Phys.: Condens. Matter 20, 434203 (2008).
\bibitem{schmid2008} H. Schmid, J. Phys.: Condens. Matter 20, 434201 (2008). 
\bibitem{wang2003} J. Wang, J. B. Neaton, H. Zheng1, V. Nagarajan, S. B. Ogale, B. Liu, D. Viehland, V. Vaithyanathan, D. G. Schlom, U. V. Waghmare, N. A. Spaldin, K. M. Rabe, M. Wuttig, and R. Ramesh, Science 299, 1719 (2003).  %Epitaxial BiFeO3 Multiferroic Thin Film Heterostructures 
\bibitem{spaldin2005} N. A. Spaldin, and M. Fiebig: The renaissance of magnetoelectric multiferroics, Science 309, 391 (2005).  %The Renaissance of Magnetoelectric Multiferroics
\bibitem{eerenstein2006} W. Eerenstein, N. D. Mathur, and J. F. Scott, Nature (London) 442, 759 (2006). %Multiferroic and magnetoelectric materials
\bibitem{cheong2007} S.-W. Cheong and M. Mostovoy, Nat. Mater. 6, 13 (2007). %Multiferroics: a magnetic twist for ferroelectricity
\bibitem{van2007} B. B. Van Aken, J. P. Rivera, H. Schmid, and M. Fiebig, Nature 449, 702 (2007).
\bibitem{ressouche2010} E. Ressouche, M. Loire, V. Simonet, R. Ballou, A. Stunault, and A. Wildes, Phys. Rev. B 82, 100408 (2010).
\bibitem{ding2016a} L. Ding, C. V. Colin, C. Darie, J. Robert, F. Gay, and P. Bordet, Phys. Rev. B 93, 064423 (2016).
\bibitem{baum2013} M. Baum, K. Schmalzl, P. Steffens, A. Hiess, L. P. Regnault, M. Meven, P. Becker, L. Bohatý, and M. Braden, Phys. Rev. B 88, 024414 (2013).
\bibitem{Fischer1972} E. Fischer, G. Gorodetsky, and R. M. Hornreich, Solid State Commu. 10, 1127 (1972).
\bibitem{kolo2011} T. Kolodiazhnyi, H. Sakurai, and N. Vittayakorn, Appl. Phys. Lett. 99, 132906 (2011). %Spin-flop driven magnetodielectric effect in Co4Nb2O9
\bibitem{fang2014} Y. Fang et al., Sci. Rep. 4 3860 (2014). %Large magnetoelectric coupling in Co4Nb2O9
\bibitem{khanh2016} N. D. Khanh, N. Abe, H. Sagayama, A. Nakao, T. Hanashima, R. Kiyanagi, Y. Tokunaga, and T. Arima, Phys. Rev. B 93, 075117 (2016).
\bibitem{yin2016} L. H. Yin, Y. M. Zou, J. Yang, J. M. Dai, W. H. Song, X. B. Zhu, and Y. P. Sun, Appl. Phys. Lett. 109, 032905 (2016).
\bibitem{bertaut1961} E. F. Bertaut, L.Corliss, F. Forrat, R. Aleonard, and R. Pauthenet, J. Phys. Chem. Solids 21, 234 (1961). %R. Etude de niobates et tantalates de metaux de transition bivalents
\bibitem{Dzyaloshinskii1959} I.E. Dzyaloshinskii, Zh. Exp. Teor. Fiz. 37, 881 (1959) [Soviet Phys. JETP 10, 628 (1960)]
\bibitem{astrov1960} D. N. Astrov, Zh. Exp. Teor. Fiz. 38, 984 (1960) [Soviet Phys. JETP 11, 708 (1960)]
\bibitem{kimura2013} A. Iyama and T. Kimura, Phys. Rev. B 87, 180408(R) (2013)
\bibitem{fiebig1994} M. Fiebig, D. Frohlich, B. B. Krichevtsov, and R. V. Pisarev, Phys. Rev. Lett. 73, 2127 (1994).
\bibitem{mcgurie1956} T. R. McGuire, E. J. Scott, and F. H. Grannis, Phys. Rev. 102, 1000 (1956).
\bibitem{schwarz2010} B. Schwarz, D. Kraft, R. Theissmann, and H. Ehrenberg, J. Magn. Magn. Mater. 322, L1 (2010).
\bibitem{deng2018} G. Deng, Y. Cao, W. Ren, S. Cao, A. J. Studer, N. Gauthier, M. Kenzelmann, G. Davidson, K. C. Rule, J. S. Gardner, P. Imperia, C. Ulrich, and G. J. McIntyre, Phys. Rev. B 97, 085154 (2018).
\bibitem{khanh2017} N. D. Khanh, N. Abe, S. Kimura, Y. Tokunaga, and T. Arima, Phys. Rev. B 96, 094434 (2017). 
\bibitem{rigaku} Rigaku, (2005) CrystalClear. Rigaku Corporation, Tokyo, Japan.
\bibitem{higashi2000} T. Higashi, ABSCOR (2000). Rigaku Corporation, Tokyo, Japan.
\bibitem{fullprof} J. Rodriguez-Carvajal, Physica B 192 55 (1993).
\bibitem{lee2014} M. Lee, J. Hwang, E. S. Choi, J. Ma, C. R. Dela Cruz, M. Zhu, X. Ke, Z. L. Dun, and H. D. Zhou, Phys. Rev. B 89, 104420 (2014). 
\bibitem{hb3a} B. C. Chakoumakos, H. Cao, F. Ye, A. D. Stoica, M. Popovici, M. Sundaram, W. Zhou, J. S. Hicks, G. W. Lynn, and R. A. Riedel, J. Appl. Cryst. 44, 655 (2011).
\bibitem{solovyev2016} I. V. Solovyev and T. V. Kolodiazhnyi, Phys. Rev. B 94, 094427 (2016).
\bibitem{shirane1965} L. M. Corliss, J. M. Hastings, R. Nathans, and G. Shirane, J. Appl. Phys. 36, 1099 (1965). %Magnetic Structure of Cr2O3
\bibitem{bilbao} J. M. Perez-Mato, S. V. Gallego, E. S. Tasci, L. Elcoro, G. de la Flor, and M. I. Aroyo, Annu. Rev. Mater. Res. 45, 217 (2015).
\bibitem{hutanu2012} V. Hutanu, A.Sazonov, M. Meven, H. Murakawa, Y. Tokura, S. Bordacs, I. Kezsmarki, and B. Nafradi, Phys. Rev. B 86, 104401 (2012). 
\bibitem{ding2016b} L. Ding, C. V. Colin, C. Darie, and P. Bordet, J. Mater. Chem. C 4, 4236, (2016).
\bibitem{zou2014} T. Zou, Z. Dun, H. Cao, M. Zhu, D. Coulter, H. Zhou, and X. Ke, Appl. Phys. Lett. 105, 052906 (2014).
\bibitem{ngo2015} T. N. M. Ngo, U. Adem, and T. T. M. Palstra, Appl. Phys. Lett. 106, 152904 (2015).
\bibitem{DOE} {https://www.energy.gov/downloads/doe-public-access-plan}.


\end{thebibliography}
\end{document}